\documentclass{article}
\usepackage[utf8]{inputenc}
\usepackage{titlesec}
\usepackage{graphicx}
\usepackage{asymptote}
\usepackage{xfrac}
\usepackage{float}
\usepackage{array}
\usepackage[ruled, vlined]{algorithm2e}
\usepackage{mathtools}
\usepackage{amsmath}
\usepackage{amsfonts}
\usepackage{amsthm}
\usepackage{amssymb}
\usepackage{pgfplots}
\usepackage[margin=1in]{geometry}
\usepackage{indentfirst}
\usepackage[style=numeric]{biblatex}
\titleformat{\section}[block]{\Large\bfseries\filcenter}{}{0em}{}
\titleformat{\subsection}[block]{\large\bfseries}{}{0em}{}
\DeclareMathOperator*{\argmin}{arg\,min}
\DeclareMathOperator*{\argmax}{arg\,max}
\pgfplotsset{width=10cm,compat=1.9}
\title{\vspace{-2em}\hrule\vspace{1em}ELMOPP: An Application of Graph Theory and\\ Machine Learning to Traffic Light Coordination\vspace{1em}\hrule}
\author{Fareed Sheriff}
\date{}
\begin{document}
\maketitle
\begin{abstract}
Traffic light management is a broad subject with various papers published that put forth algorithms to efficiently manage traffic using traffic lights. Two such algorithms are the OAF (oldest arrival first) and ITLC (intelligent traffic light controller) algorithms. However, many traffic light algorithms do not consider future traffic flow and therefore cannot mitigate traffic in such a way as to reduce future traffic in the present. This paper presents the Edge Load Management and Optimization through Pseudo-flow Prediction (ELMOPP) algorithm, which aims to solve problems detailed in previous algorithms; through machine learning with nested long short-term memory (NLSTM) modules and graph theory, the algorithm attempts to predict the near future using past data and traffic patterns to inform its real-time decisions and better mitigate traffic by predicting future traffic flow based on past flow and using those predictions to both maximize present traffic flow and decrease future traffic congestion. Furthermore, while ITLC and OAF require the use of GPS transponders; and GPS, speed sensors, and radio, respectively, ELMOPP only uses traffic light camera footage, something that is almost always readily available in contrast to GPS and speed sensors. ELMOPP was tested against the ITLC and OAF traffic management algorithms using a simulation modeled after the one presented in the ITLC paper, a single-intersection simulation, and the collected data supports the conclusion that ELMOPP statistically significantly outperforms both algorithms in throughput rate, a measure of how many vehicles are able to exit inroads every second.
\end{abstract}
\section{Introduction}
\subsection{Background}
Traffic lights used specifically for road systems have been around since the 19th century with the installation of a gas-lit traffic light in London. This traffic light was a single light that controlled horse-drawn carriage traffic and was prone to explosions. This was soon followed by the first electric light, installed in Ohio in 1914; this was also a standalone traffic light. The first network of traffic lights was implemented in Salt Lake City, Utah in 1917 as a collection of six traffic lights controlled through a manual switch. The purpose of traffic lights was to control traffic to prevent jams and decrease the risk of accidents. This is a heavy task to conduct manually because it requires that humans decide the optimal or even just an adequate configuration of traffic light timings to minimize traffic congestion, which becomes increasingly more difficult to manage as the number of traffic lights increase \cite{brefhist}.
\par As a result, the synchronization of traffic lights has largely been relegated to computer systems that take in real-time data and attempt to optimally coordinate traffic lights to reduce traffic congestion.
\subsection{Related Works}
\par One of the most commonly-used systems for this task is the Sydney Coordinated Adaptive Traffic System (SCATS), which is a combination of myriad technologies to produce real-time metrics used in tracking and managing traffic created around 1970 and used around the world. These include specialized controllers and in-road sensors. Due to the sheer volume of data and equipment necessary to set up and run SCATS, it is extremely costly and requires significant physical modifications to existing road systems to function at full capacity \cite{scatsbroc}.
\par Another early traffic system created for use in the United Kingdom is the Split Cycle and Offset Optimisation Technique (SCOOT) \cite{BRETHERTON1990237}. SCOOT makes small adjustments to traffic signal timings to increase the flow of traffic by reconciling different traffic paths - rather than drastically altering the flow of traffic, SCOOT makes small changes in an attempt to increase traffic flow in the long-term. Rather than using signal plans, SCOOT collects data in real-time using detectors installed into roads and calculates the link profile units (a combination of flow and vehicle occupancy) at each detector. SCOOT tracks the link profile units over time to produce a cyclic flow profile, which it uses to coordinate traffic across regions of the SCOOT traffic network.
\par A more recent traffic management system uses the oldest arrival-first algorithm (OAF) in conjunction with vehicular ad-hoc networks (VANETs) \cite{7119717}. It requires that all vehicles be equipped with some form of GPS to identify vehicle location at all times, speed sensors, and wireless radio. The combined use of these three devices for every vehicle forms the VANET. Groups of vehicles approaching intersections are divided into platoons and are sorted by job urgency - hence the name "oldest arrival-first". The OAF algorithm uses vehicle-specific data to schedule intersection traffic and minimize traffic delays.
\par Another algorithm that requires significantly less resources with promising results was detailed in \textit{An Intelligent Traffic Light Scheduling Algorithm Through VANETs} \cite{6927714} - the Intelligent Traffic Light Controlling algorithm (ITLC). ITLC uses the same equipment as the OAF with VANETs paper as they both use VANETs as the backbone. One key difference between ITLC and OAF, however, is that ITLC focuses on decreasing vehicle wait times rather than clearing jobs by age.
\par Over the history of traffic algorithms, it may be seen that algorithms attempt to use fewer resources and less-costly methods of data collection while also retaining the ability to keep traffic flowing. SCATS and SCOOT, both created near the end of the 20th century, have been in use for decades but the physical modifications to road systems pose a few problems that may have been acceptable in previous decades, but are not today. Although SCATS has shown convincing results for its efficacy, it is often too costly and time-consuming for many locales to install and use.
\par Recent algorithms, such as the OAF and ITLC algorithms, have improved upon previous algorithms by collecting data in a less-costly and more-efficient way - the advent of wireless networks and the Internet has allowed GPS systems to become nearly ubiquitous, which the two algorithms use to track vehicle metrics in real-time.
\par Similarly, current algorithms have focused on using machine and deep learning for vehicle traffic, such as through classifying the severity of motorcycle crashes \cite{pmid30947250} and analyzing traffic between network nodes \cite{STAUML}. While much research and attention has been focused on optimizing real-time traffic under the assumption that optimizing real-time traffic over a period of time provides for the most efficient possible flow of traffic, that assumption is not necessarily sound. Just as in a game of chess one may intentionally lose a piece in the present in order to implement a tactic that yields significant gains in future moves, it is reasonable to consider traffic optimization as a task most efficiently conducted when searching for a global optimum over a period of time rather than a local optimum in the present. In other words, the experimenter assumes that optimizing traffic by considering both past, present, and future traffic could yield more efficient traffic flow than simply optimizing traffic using only real-time data.
\section{Methodology}
\subsection{Problem Statement/Task Definition}
The algorithm this paper presents, Edge Load Management and Optimization through Pseudo-flow Prediction (ELMOPP), aims to solve the problems detailed in previous algorithms and systems --- through machine learning with nested long short-term memory modules (NLSTM), the algorithm attempts to predict the near future using past data and traffic patterns to inform its real-time decisions and better mitigate traffic. The algorithm is also easily scalable to larger road systems.
\par Furthermore, the algorithm only requires use of nearly ubiquitous traffic cameras to obtain all of the information it needs as not only are traffic cameras common on traffic lights, but if there were no traffic camera on a traffic light it would cost very little to install one. Although it was not possible in the past to classify the number of vehicles from traffic cameras, the advent of deep learning and image recognition has made efficient and accurate data collection not only feasible but also a reality. In fact, a method of counting vehicles made specifically for developing countries, but implementable anywhere, was detailed in the paper \textit{A video-based real-time adaptive vehicle-counting system for urban roads} \cite{10.1371/journal.pone.0186098}. The paper used cameras positioned to give them vantage points comparable to traffic cameras with similar quality as well and reported an accuracy above 99\% in nearly all tested scenarios, including different weather patterns. Note that the specific system used to collect data does not matter as long as the accuracy is high - the previous paper was simply cited as an example. As a result, the algorithm and associated system cost very little to install and implement, far less than most other traffic management systems, which require access to far more equipment and computing power.
\par The na{\"i}ve algorithm seeks to improve traffic flow by treating intersections as a set of possible traffic light configurations and choosing the best short-term goal at each intersection in conjunction with all of the vertex's neighbours, which it then applies to every vertex in the road network. In this way, it reaches a local-global optimum because each vertex's decision is based on the states of its neighborhood, which allows the algorithm to approximate a global optimum. The hypothesis this paper sought to test was "if the ELMOPP algorithm is tested \textit{in silico} against the ITLC and OAF traffic management algorithms, ELMOPP will exhibit a statistically significantly higher throughput rate than either algorithm."
\subsection{Algorithm}
\par A road system may be modeled as the induced directed graph $G=(V, E)$ of the road network where $V$ is the vertex set of the digraph representing all intersections of the systems while $E$ is the directed edge set contain all directed roads connecting each intersection.
\begin{figure}[H]
    \centering
    \includegraphics[width=7cm]{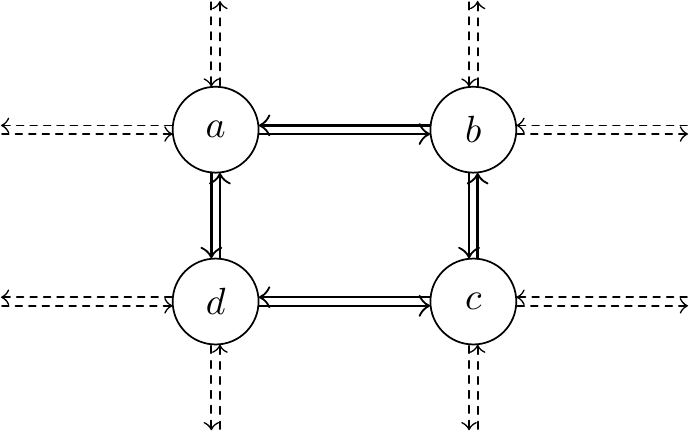}
    \caption{A sample induced digraph $G_{\alpha}$ of a road system}
\end{figure}
Note that the dashed edges connected to each vertex represent roads that are only connected to one intersection/vertex; as a result, these directed edges are termed "pseudo-diedges" as they do not fit the traditional definition of a directed edge but they are roads nonetheless. Pseudo-diedges are not included as part of the induced digraph of a road system and are drawn as a formality: they are not directly considered in calculations and are only indirectly considered through in-flow predictions of directed edges to which these pseudo-diedges form a path. Also note that the "edges" referenced prior to this point were considered edges that mapped to a single road. From this point onward, edges will be considered in both the context of roads and lanes within those roads. In this way, an edge may contain multiple edges termed subedges that map to road lanes.
\par The adjacency matrix of the induced digraph contains all vertex-vertex connections. This paper follows the row-tail, column-head adjacency matrix convention, where each row represents vertices marking the start of a diedge and each column represents vertices marking the head of a diedge. For example, the adjacency matrix $A_{\alpha}$ of the graph in Figure 1 is 
\begin{equation*}
    A_{\alpha}=\begin{bmatrix}0&1&0&1\\1&0&1&0\\0&1&0&1\\1&0&1&0\end{bmatrix}
\end{equation*}
while the vertex set $V_{\alpha}$ is $\{a, b, c, d\}$ and the directed edge set $E_{\alpha}$ is $\{(a, b), (b, a), (a, d), (d, a), (b, c),$ $(c, b),$ $(c, d), (d, c)\}$. However, because each road has four possible directions it can take (forward, left + U-turn, and right), every edge is represented as a 3-vector containing the partitioning of each road into its lanes so that every component of the 3-vector is a subedge, where the first component represents the number of vehicles in the left lane(s), the second component represents the number of vehicles in the middle lane(s), and the third component represents the number of vehicles in the right lane(s). This turns the adjacency matrix of the induced digraph into an adjacency tensor of order three. Therefore, a possible adjacency tensor of $G_{\alpha}$ could be
\begin{equation*}
    A_{\alpha}=
    \begin{bmatrix}
        \begin{bmatrix}0&0&0\end{bmatrix}&\begin{bmatrix}0&1&1\end{bmatrix}&\begin{bmatrix}0&0&0\end{bmatrix}&\begin{bmatrix}1&1&1\end{bmatrix}\\
        \begin{bmatrix}0&0&1\end{bmatrix}&\begin{bmatrix}0&0&0\end{bmatrix}&\begin{bmatrix}1&1&1\end{bmatrix}&\begin{bmatrix}0&0&0\end{bmatrix}\\
        \begin{bmatrix}0&0&0\end{bmatrix}&\begin{bmatrix}1&1&1\end{bmatrix}&\begin{bmatrix}0&0&0\end{bmatrix}&\begin{bmatrix}1&1&1\end{bmatrix}\\
        \begin{bmatrix}1&1&1\end{bmatrix}&\begin{bmatrix}0&0&0\end{bmatrix}&\begin{bmatrix}1&1&1\end{bmatrix}&\begin{bmatrix}0&0&0\end{bmatrix}
    \end{bmatrix}
\end{equation*}
\par Let $C^{i\times j\times k}$ be the capacity tensor that maps a maximum vehicle capacity to every edge in the adjacency tensor and a variable quantity tensor $Q^{i\times j\times k}$ that contains the number of vehicles on every road lane at time $t$. It follows that the load $L$ of $G$ equals the Hadamard quotient of the quantity and the capacity\footnote{The load of an edge is artificially set to 0 when the capacity of an edge is 0 as it is not possible for any vehicles to travel on that edge. In setting the edge load to 0 when the capacity is 0, the urgency of that edge becomes 0, preventing so-called "dead" edges from having any effect on configuration urgency.}.
\begin{equation}
    L = Q \oslash C
\end{equation}
As an example, the quantity, capacity, and load tensors for $G_{\alpha}$ could be
\begin{equation*}
    Q_{\alpha}=
    \begin{bmatrix}
        \begin{bmatrix}0&0&0\end{bmatrix}&\begin{bmatrix}0&2&3\end{bmatrix}&\begin{bmatrix}0&0&0\end{bmatrix}&\begin{bmatrix}0&6&4\end{bmatrix}\\
        \begin{bmatrix}0&0&3\end{bmatrix}&\begin{bmatrix}0&0&0\end{bmatrix}&\begin{bmatrix}2&0&1\end{bmatrix}&\begin{bmatrix}0&0&0\end{bmatrix}\\
        \begin{bmatrix}0&0&0\end{bmatrix}&\begin{bmatrix}2&1&3\end{bmatrix}&\begin{bmatrix}0&0&0\end{bmatrix}&\begin{bmatrix}4&5&4\end{bmatrix}\\
        \begin{bmatrix}7&7&7\end{bmatrix}&\begin{bmatrix}0&0&0\end{bmatrix}&\begin{bmatrix}3&2&0\end{bmatrix}&\begin{bmatrix}0&0&0\end{bmatrix}
    \end{bmatrix}
\end{equation*}
\begin{equation*}
    C_{\alpha}=
    \begin{bmatrix}
        \begin{bmatrix}0&0&0\end{bmatrix}&\begin{bmatrix}0&4&4\end{bmatrix}&\begin{bmatrix}0&0&0\end{bmatrix}&\begin{bmatrix}3&9&6\end{bmatrix}\\
        \begin{bmatrix}0&0&7\end{bmatrix}&\begin{bmatrix}0&0&0\end{bmatrix}&\begin{bmatrix}3&2&4\end{bmatrix}&\begin{bmatrix}0&0&0\end{bmatrix}\\
        \begin{bmatrix}0&0&0\end{bmatrix}&\begin{bmatrix}3&5&3\end{bmatrix}&\begin{bmatrix}0&0&0\end{bmatrix}&\begin{bmatrix}7&8&9\end{bmatrix}\\
        \begin{bmatrix}9&8&7\end{bmatrix}&\begin{bmatrix}0&0&0\end{bmatrix}&\begin{bmatrix}4&5&5\end{bmatrix}&\begin{bmatrix}0&0&0\end{bmatrix}
    \end{bmatrix}
\end{equation*}
\begin{equation*}
    L_{\alpha}=
    \begin{bmatrix}
        \begin{bmatrix}0&0&0\end{bmatrix}&\begin{bmatrix}0&\sfrac{1}{2}&\sfrac{3}{4}\end{bmatrix}&\begin{bmatrix}0&0&0\end{bmatrix}&\begin{bmatrix}0&\sfrac{2}{3}&\sfrac{2}{3}\end{bmatrix}\\
        \begin{bmatrix}0&0&\sfrac{3}{7}\end{bmatrix}&\begin{bmatrix}0&0&0\end{bmatrix}&\begin{bmatrix}\sfrac{2}{3}&0&\sfrac{1}{4}\end{bmatrix}&\begin{bmatrix}0&0&0\end{bmatrix}\\
        \begin{bmatrix}0&0&0\end{bmatrix}&\begin{bmatrix}\sfrac{2}{3}&\sfrac{1}{5}&1\end{bmatrix}&\begin{bmatrix}0&0&0\end{bmatrix}&\begin{bmatrix}\sfrac{4}{7}&\sfrac{5}{8}&\sfrac{4}{9}\end{bmatrix}\\
        \begin{bmatrix}\sfrac{7}{9}&\sfrac{7}{8}&1\end{bmatrix}&\begin{bmatrix}0&0&0\end{bmatrix}&\begin{bmatrix}\sfrac{3}{4}&\sfrac{2}{5}&0\end{bmatrix}&\begin{bmatrix}0&0&0\end{bmatrix}
    \end{bmatrix}
\end{equation*}
\par When a traffic light is active, it is often the case that another traffic light at the same intersection could also be active. For example, at a four-way intersection the left lane green light may be active for two antiparallel inroads. More specifically, the only two possible non-intersecting configurations for a pair of antiparallel inroads to an intersection are both left lanes active and both middle and right lanes active. This may be seen in US road system traffic light coordination. A traffic light configuration is defined to be a set of subedges directed toward a common vertex so that none of the subedges' traffic streams intersect. The activation of a traffic light configuration is defined to be the activation of each traffic light corresponding to its associated subedge in a configuration. There are eight such configurations for a four-way intersection: two for two antiparallel inroads,  two for the other pair of antiparallel inroads, and four for each inroad. The configuration set for a vertex $v$ is symbolized $C(v)$ and contains all valid configurations for a vertex while a specific configuration on $v$ is symbolized $C_v$, where $C_{v_n} \in C(v)$.\footnote{Note that, although four-road intersections are most often mentioned in this paper, intersections of any possible number occur in the real world. Four-road intersections are used as examples here simply because they are common, it is easy for audiences to understand the ELMOPP algorithm through easily-relatable examples, and because the simulation used to test the algorithm uses a four-road intersection in keeping with the ITLC paper \cite{6927714}. However, the idea of a configuration is valid for all types of intersections. All that must be done is define the configuration set as the set of all possible configurations.}
\subsubsection{Na{\"i}ve Algorithm}
\par Define the urgency of subedge $e_n$ as a function of the load of the subedge and the time $T$ since the last activation of a subedge so that the urgency of subedge $e_n$, $U(e_n)$, is defined to be
\begin{equation}
    U(e_n) = \frac{L(e_n)}{e^{1 - \frac{T(e_n)}{t_{max}}}} = L(e_n)e^{\frac{T(e_n)}{t_{max}} - 1}
\end{equation}
and the urgency of orientation $C_{v_n}$, $U(C_{v_n})$, is defined to be
\begin{equation}
    U(C_{v_n}) = \sum_{e_m\in C_{v_n}}{U(e_m)}
\end{equation}
where $L(e_n)$ is the load of subedge $e_n$, $T(e_n)$ is the time passed since $e_n$ was last activated, and $t_{max}$ is the maximum legal activation time for a traffic light. The urgency of an edge configuration is defined as the sum of the urgencies of all elements of the edge's configuration set. Equation (3) simply sums together the urgencies of every subedge in a configuration. Equation (2) is an exponential function meant to prioritize edges that have not been activated for longer periods of time as well as edges that have high loads. The formula is similar to the logistic equation, save that it is exponential due to the removal of the unit addition in the denominator of the first fraction in (2). The minimum value an edge's urgency may take is $\frac{L(e_n)}{\mbox{e}}$, while there is no maximum. However, the urgency of an edge when the time since the last activation equals $t_{max}$ is $L(e_n)$, or just the load of an edge itself.
\par The algorithm simply chooses the configuration with the highest urgency for vertex $v$, $C_{v+}$
\begin{equation}
    C_{v+} = \max_{C_v\in C(v)}U(C_v)
\end{equation}
This configuration is held until
\begin{equation}
    \max_{C_v\in C(v)\backslash C_{v+}}U(C_v) \geq U(C_{v+})
\end{equation}
for
\begin{equation}
    t\in (t_{min}, t_{max})
\end{equation}
otherwise, 
\begin{equation}
    t = \argmin_{p \in \{t_{min}, t_{max}\}}{|p - t|}
\end{equation}
where $t_{min}$ is the minimum legal activation time for a traffic light. This is so that the configuration is activated until the urgency directly before the configuration was activated equals the urgency of another configuration directly before it was activated, while keeping in mind minimum and maximum green light length rules. The algorithm is conducted in real-time, so there is no need for a model of unload speed; rather, real-time data is used and the rules above are used when monitoring load. In fact, because the algorithm is conducted in real-time, the modelling of out-flow may be safely ignored in its entirety as no modelling is required when real-time data is available. As a result, it suffices to only consider edge \& subedge in-flow in the algorithm.
\subsubsection{Cumulative Algorithm}
\par The previously-detailed algorithm is a na{\"i}ve algorithm similar to those of various systems used to coordinate traffic lights in that these algorithms fail to consider future traffic and how to best prevent traffic density from increasing through future predictions. As may be obvious, the intersection-road induced graph fails to consider the number of vehicles at places like department stores or office buildings because
\begin{enumerate}
    \item taking those places into account would result in an overly complex graph and "road" system and
    \item violating the conservation of graph flow allows for generalized predictions to be made by relegating the entrances and exits of vehicles into and from buildings to negative and positive edge flow. 
\end{enumerate}
\par Therefore, rather than calling the movement of vehicles between edges flow, which it inherently is not due to the lack of conservation of graph flow, flow shall hereon be considered to be pseudo-flow, a construct defined to be equivalent to flow in all respects save for obeying the conservation of flow. It is then possible to create a vector each of whose elements correspond to the flow of a certain edge over time. Furthermore, the closer in time the future flow of an edge is to the current time of an edge, the greater the effect the future flow will play in the configuration decision at an intersection. Furthermore, the effect of future edge flows may be modelled using a bounded sine function whose maximum occurs at the current time and minimum at the maximum configuration activation time. The cumulative urgency $U_c$ of a given path may be expressed as the convolution integral transform between current and future urgencies and a negatively-sloped line whose area from the origin to the $x$-intercept equals 1, effectively modeling a triangle with legs on the $x$- and $y$-axes and base $t_{max}$. With a bit of elementary geometry, it is seen that the height of the triangle must be 
\begin{equation}
    bh = ht_{max} = 2
\end{equation}
\begin{equation}
    h = \frac{2}{t_{max}}
\end{equation}
Therefore, the equation of the line is
\begin{equation}
    y = h - \frac{h}{b}x = 2\left(\frac{1}{t_{max}} - \frac{x}{t_{max}^2}\right)
\end{equation}
The cumulative urgency convolution then becomes
\begin{equation}
    U_c(e_n)=(\triangle*\, U)(t)=2\int_0^{t_{max}}{U_t(e_n)\left(\frac{1}{t_{max}} - \frac{x}{t_{max}^2}\right)}\, dt
\end{equation}
where
\begin{equation}
    U_0(e_n) = U(e_n)
\end{equation}
and
\begin{equation}
    U_t(e_n) = L(e_{n_t})e^{1-\frac{T(e_n)}{t_{max}}}
\end{equation}
where $e_{n_t}$ is the subedge $e_n$ at time $t$.
\par Future flow is predicted using a recurrent neural network (RNN) due to the fact that the current in-flow of a directed edge both is roughly periodic in nature and is a product of previous in-flows. For example, it is to be expected that certain roads will see above-average in-flows during rush hour. Similarly, in-flows cannot stay high for long periods of time if the in-flows of other roads have been low because the low in-flows of other roads means that the road with high in-flow is getting saturated and is approaching capacity. The RNN used is, to decrease computation times, simplify calculations, and take into account the inherently discrete nature of traffic light timing, a single triply-nested long short-term memory (NLSTM) unit \cite{pmlr-v77-moniz17a}. Nested LSTMs present a solution to this problem by replacing the long-term memory unit of an LSTM with another LSTM, thus increasing the long-term memory span of the nested LSTM. The reason behind the use of a triply-nested LSTM rather than a conventional LSTM \cite{Hoch97} is that the time-steps used in the model occur on the order of minutes, but patterns in traffic in-flow often occur on the order of days, months, or even years. One oft-noted problem with vanilla LSTMs is that although they are competent at recognizing patterns occurring over hundreds of time-steps, their long-term memory fail to remember patterns spanning thousands or more time-steps \cite{Hoch98, 7508408}. Because there are around $365*24*60*60=32,850,000$ minutes or time-steps in a year, which is close to $\log_{500}{32,850,000}\approx 2.7784\approx 3$ magnitudes greater than a single LSTM can handle, it is necessary to have at least three levels of LSTMs to properly account for patterns spanning long ranges of time.
Due to the discreteness of both the RNN and the configuration activations, the cumulative urgency integral must be transformed into a discrete convolution summation
\begin{equation}
    U_c(e_n)= (\triangle*\, U)[t] = 2\sum_{t = 0}^{t_{max}}{\left[U_t(e_n)\left(\frac{1}{t_{max}} - \frac{x}{t_{max}^2}\right)\right]}
\end{equation}
The cumulative urgency is substituted for the na{\"i}ve urgency when relegating intersection configuration so that $C_{v+}$ becomes
\begin{equation}
    C_{v+} = \max_{C_v\in C(v)}U_c(C_v)
\end{equation}
\subsubsection{Complete Algorithm}
\par The complete algorithm is as follows, where each run-through of the \textbf{while} loop represents a single timestep.\\
\begin{algorithm}[H]
    \SetAlgoLined
    \KwData{\textit{CONFSET}: configuration set for specific intersection; \textit{CONF}: configuration in \textit{CONFSET}; \textit{LENCONF}: length since configuration was last activated; \textit{CCONF}: current active configuration}
    Pretrained NLSTM is readied\;
    \For{each intersection in road network}{
        \While{traffic flows}{
            Calculate cumulative urgency for each \textit{CONF} \textbf{in} \textit{CONFSET} using NLSTM\;
            Input real-time flow information into NLSTM for training\;
            Increment \textit{LENCONF} for all \textit{CONF} \textbf{in} \textit{CONFSET} while keeping \textit{LENCONF} of \textit{CCONF} 0\;
            \eIf{cumulative urgency of \textit{CCONF} = $\displaystyle\max_{CONFSET}{\{\mbox{cumulative urgency of \textit{CONF}}\}}$}{
                do not switch configuration\;
            }{
            \textit{CCONF} = $\displaystyle\argmax_{CONFSET}{\{\mbox{cumulative urgency of \textit{CONF}}\}}$\;
            }
        }
    }
    \caption{ELMOPP algorithm}
\end{algorithm}
The prediction portion of the algorithm is governed by the NLSTM, which has trained on past traffic inflow data; the assumption is made that the more traffic data has been trained upon, the more comprehensive the patterns the NLSTM learns are. At any moment, the NLSTM is able to predict future inflows by outputting predictions when given current or predicited future inflows. It is through this method that the NLSTM is able to predict inflows tens of timesteps into the future.
\section{Evaluation}
A simulation will be conducted to test the detailed algorithm. This simulation will be of the same form as the simulation detailed in the ITLC paper \cite{6927714} so as to facilitate direct comparison between the ITLC, OAF, and ELMOPP algorithms. The simulation used in the ITLC paper was a single 4-leg traffic intersection simulation with a simulation area of 1,000 square meters. The simulation was also conducted over 2000 seconds with anywhere from 200 to 1000 vehicles and a 1.5-second start-up time (time loss due to starting vehicle) per vehicle. Likewise, the simulation used in this paper to evaluate the ELMOPP algorithm was a single 4-leg traffic intersection simulation that took place over 2000 seconds, with each second a single timestep, as well as a 1.5-second startup time for each vehicle with 200 to ,000 vehicles. The simulation graph $G_S$ representing the 4-leg traffic intersection used in the simulation is the induced digraph of the simple 4-star.
\begin{figure}[H]
    \centering
    \includegraphics[width=7cm]{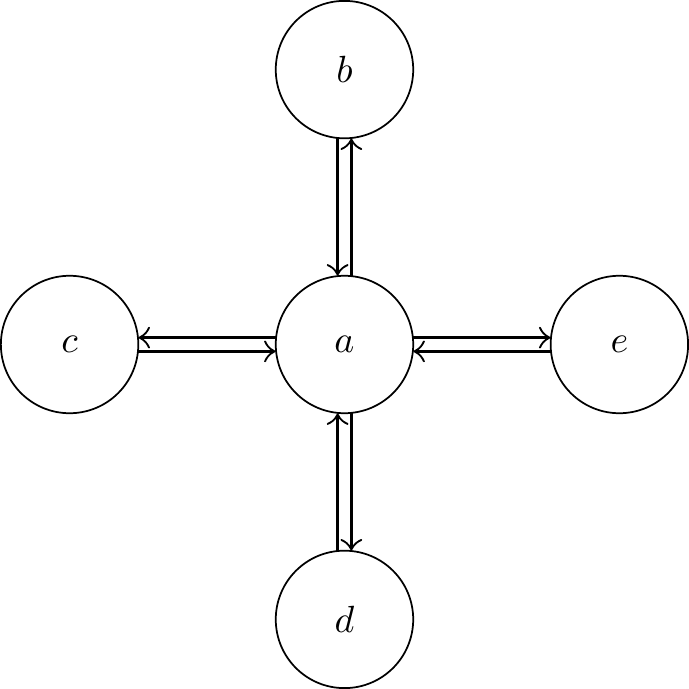}
    \caption{The induced digraph $G_{S}$ of the simple 4-star, used to simulate the algorithm}
\end{figure}
Note that because the simulation detailed in previous research \cite{6927714} only considers traffic inbound to vertex $a$, the algorithm will not account for outgoing traffic originating from $a$; in other words, the loads of all outgoing edges from $a$ are artificially set to a constant 0. Following previous work, this simulation considers the length of each time-step to be a second; however, unlike the simulation, this simulation does not randomly generate per-second inflows because road inflow in real life is not random but chaotic. Mathematically, chaotic and random flow are similar in that they both appear random, but chaos is deterministic and true randomness is non-deterministic. In that respect, while chaotic flow is used in this simulation, chaos is similar in nature to the randomness used in the ITLC paper. The simulation time of the provided simulation is 2000 seconds/time-steps while the total number of vehicles is 200 to 1,000, with a set of thirty trials conducted. Therefore, a periodic in-flow function will be enacted to better simulate real-world traffic inflow. This in-flow function is a 4-dimensional autonomous hyperchaotic system modelled off the Lorenz attractor as described in the hyperchaotic system paper \cite{Zhang2017}. That system of differential equations (see appendix) exhibits chaotic behaviour, which models the real world.
\par To conform to the simulation used, the hyperchaotic system will be normalized by dividing the values generated at each timestep by the integral of the system over the interval $[0, 2000]$, then will be scaled by a random integer over the interval $[200, 1000]$. This results in chaotic inflow with, following previous work, a total number of incoming vehicles between 200 and 1,000. The capacity of each road was not given in the original paper's simulation description; therefore, the capacity of every inroad to $a$ is set to be a constant $1,000$. Note that the value of the capacity itself may be arbitrarily positive because all that matters is that the capacities themselves are equal across all edges, as only capacity ratios are considered as can be seen from the urgency formula. The simulation previously described also contains other information, such as the area over which the simulation is conducted, variables that are extraneous to this simulation as ELMOPP does not require these variables.
\par The LSTM used in the simulation will not be a triply-nested LSTM, but will instead be a vanilla LSTM (one may consider this to be a singly-nested LSTM) because the hyperchaotic system will be timewise scaled down to produce patterns over time intervals ranging from minutes to hours due simply to the length of the simulation detailed in the ITLC paper\cite{6927714} (around 33 minutes over 2,000 single-second timesteps). In other words, the LSTM used in the simulation will be a vanilla LSTM simply because the time scale of the simulation is magnitudes shorter than the time scale of real-world traffic patterns; therefore, a triply-nested LSTM is not needed as the time scale it handles dwarfs the time scale of the simulation and is more computation-intensive than the better-suited LSTM (better-suited only for the simulation). As a result, the RNN used will not need to be able to handle large time scales. More details on the simulation are provided in the appendix.
\par The results from the simulation were collected and compared against the results from the ITLC and OAF simulations \cite{6927714}, as shown in the plot below. The data collected were plotted against the results of the ITLC and OAF algorithms.
\begin{figure}[H]
    \centering
    \begin{tikzpicture}
        \begin{axis}[
            title={Throughput vs. travelling vehicle count of all three algorithms},
            xlabel={Number of travelling vehicles},
            ylabel={Throughput (vehicles per second)},
            xmin=200, xmax=1000,
            ymin=1, ymax=2.4,
            xtick={200,400,600,800,1000},
            ytick={1,1.2,1.4,1.6,1.8,2.0,2.2, 2.4},
            legend pos=north east,
            ymajorgrids=true,
            grid style=dashed]

            \addplot[color=red!20!yellow, mark=square,]
                coordinates{(200,2.31937)(400,2.17839)(600,2.06249)(800,1.97497)(1000,1.92143)};
                \addlegendentry{ELMOPP}
            \addplot[color=red, mark=square,]
                coordinates{(200,1.692)(400,1.883)(600,1.8520)(800,1.8593)(1000,1.8844)};
                \addlegendentry{ITLC}
            \addplot[color=green, mark=square,]
                coordinates{(200,1.2174)(400,1.05736)(600,1.1540)(800,1.1057)(1000,1.07094)};
                \addlegendentry{OAF}
        \end{axis}
    \end{tikzpicture}
\end{figure}
\par Furthermore, an independent samples \textit{t}-test was conducted for each pair of algorithms. The mean of the ELMOPP algorithm is 2.09133 and the standard deviation 0.158824. The mean of the ITLC algorithm is 1.83414 with a standard deviation of 0.080730. Finally, the mean of the OAF algorithm is 1.12108 with a pooled standard deviation of 0.063498. Thirty trials were conducted for each data point provided for every algorithm. Three independent samples \textit{t}-tests were conducted for every pair of algorithms. It was assumed that the data distributions and the variances for each algorithm were different when conducting the tests, supported by the calculated standard deviations. The null hypothesis was that the means of the distributions were equal while the alternate hypothesis was that the greater mean was actually greater than the lesser mean. Therefore, the \textit{t}-test was one-sided. Finally, the degrees of freedom were set to 298 for each test as there were 150 data points for each dataset and the level of significance was set to 0.05. The results of the \textit{t}-tests are shown below.
\begin{table}[H]
    \centering
    \newcolumntype{g}{>{\centering\arraybackslash}m{4.4cm}}
    \newcolumntype{h}{>{\centering\arraybackslash}m{3.3cm}}
    \begin{tabular}{|g|h|h|h|}
        \hline \textbf{Test} & \textbf{Calculated \textit{t}-value} & \textbf{Table \textit{t}-value} & \textbf{Significance}\\\hline\hline
        \textit{t}-test: ELMOPP v. ITLC & 17.6799 & 1.6500 & Significant\\\hline
        \textit{t}-test: OAF v. ELMOPP & 69.4727 & 1.6500 & Significant\\\hline
        \textit{t}-test: ITLC v. OAF & 85.0275 & 1.6500 & Significant\\\hline
    \end{tabular}
\end{table}
At the 95\% confidence level, it is seen that the alternate hypothesis is supported for each pair of algorithms. This, coupled with the algorithm's means in order from greatest to least as ELMOPP, ITLC, and OAF, supports the hypothesis that ELMOPP exhibits a higher throughput than both ITLC and OAF. Surprisingly, the calculated \textit{t}-value for the ITLC v. OAF test is the greatest of all calculated \textit{t}-values. This is because, even though the difference of means for this test is smaller than the difference of means for the ELMOPP v. OAF, the variance for the ITLC dataset is also smaller than the variance for the ELMOPP dataset.
\section{Applications \& Further Research}
\par Applications of the novel ELMOPP algorithm are varied in scope. As a traffic management algorithm built to keep traffic flow in road systems as high as possible, ELMOPP could be used on practically any road system. ELMOPP would be especially useful in places that do not have access to GPS systems or aren't able to pay for extreme renovations to road systems as would be required by systems like SCATS or SCOOT. This includes places such as cities with low funds or towns with minimal extra resources. In fact, ELMOPP could be applied to any road system that needs a traffic management system quickly as it is straightforward to implement and requires practically no physical modifications to existing road systems. The observed increase in throughput shown by ELMOPP in comparison with ITLC and OAF is beneficial for another reason: environmental impact. Increased traffic congestion has been shown to correlate with lower ambient air quality and has even been linked to trends in increasing mortality. Hazards associated with traffic congestion have been shown to be related to travel time and rush hour length, among others \cite{Zhang2013}. As ELMOPP seems to show a higher throughput than ITLC and OAF, it is safe to say that travel time will be reduced as more vehicles at any given moment are travelling to their destination for ELMOPP than for either of the other two algorithms. Furthermore, ELMOPP was specifically created to predict and mitigate future traffic, something that has immediately-obvious implications for rush hour traffic. This further supports the application of ELMOPP to decreasing environmental impact and positively impacting people's health.
\par Further research on the subject of traffic management algorithms would likely involve different methods of approaching and describing the problem. ITLC used algebra and basic combinatorics to create what was a very loose description of road systems; ELMOPP used graph theory and linear algebra to describe and optimize traffic. There are other, potentially more scalable, methods of describing traffic flow. Further research by the researcher would likely be focused on the avenues of other methods of predicting traffic flow, such as through vanilla artificial neural networks (ANNs) or graph convolutional networks (GCNs). One specific venue for further research could be real-world testing by implementing ELMOPP for a physical road system and comparing its performance to other traffic management algorithms. Another, more comprehensive, method of further research might involve developing a suite of network tests to cover a range of situations such as alternating heavy and light traffic flow, various intersection types, traffic management algorithms applied to simulated road networks rather than individual intersections, and counting roads with varying speed limits.
\section{Conclusion}
\par This paper outlined a novel traffic management algorithm named ELMOPP that employed graph theory and nested LSTMs. The ELMOPP algorithm assumes that current traffic flow is correlated with past traffic flow and that traffic cameras are able to collect data that can be used to determine incoming traffic parameters. In contrast to real-time traffic optimization, the ELMOPP algorithm attempts to optimize traffic across time by searching for the most efficient traffic flow using machine-learned traffic patterns to predict future traffic.
\newpage
\printbibliography
\newpage
\section{Appendix}
\subsection*{Hyperchaotic System Inflow Simulations}
The hyperchaotic system governing inflow dynamics for the simulation was
\begin{equation}
    (x, y, z, w)\coloneqq\begin{cases}
        \frac{dx}{dt} = a(y-x)-ew,\\
        \frac{dy}{dt} = xz-hy,\\
        \frac{dz}{dt} = b - xy - cz,\\
        \frac{dw}{dt} = ky - dw
    \end{cases}
\end{equation}
where "$a$, $b$, $c$, $d$, $e$, $h$ are positive parameters of system" \cite{Zhang2017} and with the specific attractor a = 5, b = 20, c = 1, d = 0.1, e = 20.6, h = 1, and k = 0.1. As stated by the paper, the largest Lyapunov exponent of the attractor above is 0.24. As a result, the Lyapunov time of the system is $\sfrac{1}{0.24}\approx 4.167$. Because each system time-step is 0.01, the Lyapunov time expressed in time-steps is $4.167/0.01\approx 417$, which amounts to $\sfrac{417}{60}\approx 6.95$ minutes, far greater than any reasonable maximum traffic light activation time. As a result, the cumulative urgency formula is not expected to significantly diverge from the true chaotic distribution as the Lyapunov time is far greater than any expected result.
\par The system was solved by using the Runge-Kutta family's Euler method with a step size of 0.01 over $[0, 2000]$ by treating the system as the derivative of a vector-valued function.
\par Euler's method allows for discrete approximations of differential equations. It states that for function $y$ and its derivative $y'$,
\begin{equation}
    \mathbf{h}_{n+1} = \mathbf{h}_n + \mathbf{h}'(t_n)\Delta_n
\end{equation}
This may be extended to vector-valued functions, such as the previously-described hyperchaotic system \cite{Zhang2017}, where
\begin{equation}
    \mathbf{h} = \begin{bmatrix}x\\y\\z\\w\end{bmatrix},\qquad
    \mathbf{h'} = \begin{bmatrix}a(y-x)-ew\\xz-hy\\b - xy - cz\\ky - dw\end{bmatrix}
\end{equation}
and $\Delta_n = 0.01$.
\par The system was normalized by dividing the system by the L1 norm of its integral over $[0, 2000]$, then scaling it up by a factor of 800. This effectively modeled chaotic vehicle inflow so that the total inflow over the 2,000 timesteps was 800 vehicles. The integral of the solution was calculated by summing all of the data points calculated through Euler.
\par Because the step size was 0.01 and the stiffness ratio (ignoring the eigenvalue of 0) was $-\sfrac{7.56}{0.23}\approx-32.8695$, which is not significantly less than -1, it is safe to say that the system is non-stiff for the estimation technique used. 
\subsection*{LSTM Prediction Model}
\par The LSTM prediction model trained on 10,000 data points. The data was generated by choosing a random real vector $v\in \mathrm{R}^4$ so that $v$ is a vector randomly sampled from the 4-dimensional hypercube whose boundaries are defined by the fourfold Cartesian product $\{\{0, 1\}\times \{0, 1\} \times \{0, 1\} \times \{0, 1\}\}$. The LSTM's training data was split 80-20 as traditionally split when training and testing data for machine learning models. The first 8,000 data points were used only training while the last 2,000 points were used for simulation. Every data point in the last 2,000 points was trained pointwise on the LSTM as a single-sample batch.
\par The LSTM itself consisted of 16 units (determined empirically through tests over different ranges of units to be the optimal number of units to facilitate low errors while preventing overfitting for the specific attractor used) and was trained over 100 epochs with a batch size of 10 samples using the Adam optimizer with a mean squared error loss function.
\subsection*{Traffic Intersection Simulation}
\par Few details were given on the ns2 map used in the ITLC paper \cite{6927714}; as a result, some assumptions were made about the simulation. The turning speed was set at a constant 10 meters per second, chosen from the thorough analysis done in \textit{MODELING SPEED PROFILES OF TURNING VEHICLES AT SIGNALIZED INTERSECTIONS} \cite{modprof}. Furthermore, the outflow rate ($q$) was calculated using a numerical Greenshield model \cite{transsys} to be
\begin{equation}
    q = kv_f\left(1-\frac{k}{k_j}\right)
\end{equation}
where $k$ is the density, $v_f$ is the free-flow speed, and $k_j$ is the jam density. The capacity of each street was set to 1,000 vehicles. Each inroad was initialized to a load of 200 vehicles with the total vehicle inflow over the 2,000-second testing interval being 800, with a resulting 4,000 vehicles total and an average of 1,000 vehicles per inroad, only accomplished if all 4,000 vehicles flow out over the 2,000-second interval. The capacity for each inroad was set to 1,000 with a jam density of 1.
\end{document}